\documentclass[12pt,preprint]{aastex}
\usepackage{epsfig,emulateapj5,apjfonts}
\newcommand       \Angstrom     {\,{\rm \AA}}

\newcommand       \s        {\,{\rm s}}

\newcommand       \yrs      {\,{\rm yrs}}

\newcommand       \simlt        {\lesssim}
\newcommand       \simgt        {\gtrsim}
\newcommand       \gtsim        {\gtrsim}

\newcommand       \mum          {\,{\rm \mu m}}

\newcommand       \simali   {\sim\,}
\newcommand       \Alambda  {A_\lambda}
\newcommand       \AV       {A_V}
\newcommand       \Fnu      {F_\nu}
\newcommand       \Fo       {F_{\rm o}}
\newcommand       \dof      {{\rm d.o.f.}}
\newcommand       \magni    {\,{\rm mag}}
\def	\beq	{\begin{equation}}
\def	\eeq	{\end{equation}}
\def	\beqa	{\begin{eqnarray}}
\def	\eeqa	{\end{eqnarray}}
\newcommand{\figwidth}{6.0in}

\shorttitle{Dust Extinction of GRB Host Galaxies}

\begin{document}
\title{
On Dust Extinction of Gamma-ray Burst Host Galaxies
     }
\author{Aigen Li\altaffilmark{1},
        S.~L. Liang\altaffilmark{1},
        D.~A. Kann\altaffilmark{2},
        D.~M. Wei\altaffilmark{3},
        S. Klose\altaffilmark{2},
        and Y.~J. Wang\altaffilmark{4}}
\altaffiltext{1}{Department of Physics and Astronomy, University of
                 Missouri, Columbia, MO 65211; {\sf lia@missouri.edu}}
\altaffiltext{2}{Th\"{u}ringer Landessternwarte Tautenburg, 
                 D-07778 Tautenburg, Germany}
\altaffiltext{3}{Purple Mountain Observatory, Chinese Academy of
                 Sciences, Nanjing 210008, China}

\altaffiltext{4}{Department of Physics, Hunan Normal University,
                 Changsha 410071, China} 

\begin{abstract}
Although it is well recognized that gamma-ray burst (GRB) 
afterglows are obscured and reddened by dust in their host 
galaxies, the wavelength-dependence and quantity of dust 
extinction are still poorly known. 
Current studies on this mostly rely on fitting
the afterglow spectral energy distributions (SEDs)
with template extinction models.
The inferred extinction (both quantity and 
wavelength-dependence) and dust-to-gas ratios
are often in disagreement with that obtained from 
dust depletion and X-ray spectroscopy studies.
We argue that this discrepancy could result from
the prior assumption of a template extinction law.
We propose an analytical formula to approximate
the GRB host extinction law.
With the template extinction laws self-contained,
and the capability of revealing extinction laws 
differing from the conventional ones, 
it is shown that this is a powerful approach 
in modeling the afterglow SEDs to derive GRB host extinction.  
\end{abstract}


\section{Introduction\label{sec:intro}}
In addition to the Galactic foreground extinction,
GRBs and their afterglows are 
subject to extinction caused by the dust within 
their host galaxies. Evidence for this includes ---
\begin{itemize}
\item ``{\it Dark bursts}'' -- an appreciable fraction
      of GRBs with X-ray and/or radio afterglows
      lack an optical afterglow 
      (Jakobsson et al.\ 2004).\footnote{%
        Prior to the launch of {\it Swift}, 
        nearly $\simali$60\% of the X-ray afterglows
        reportedly had no optical counterparts.
        Despite rapid and deep searches in the {\it Swift} era, 
        it was found that $\simali$1/3 GRBs with bright X-ray 
        afterglows remain undetected at optical wavelengths
        (Fiore et al.\ 2007, Schady et al.\ 2007).
       }
      A natural explanation for dark bursts is that they
      lie behind significant obscuring dust columns
      in their host galaxies
      which effectively suppresses the optical light 
      [although some dark bursts may be intrinsically 
        faint or occur at high redshifts (say, $z\gtsim 5$)
        where the Ly$\alpha$ break has moved through 
        the optical bands, leading to absorption of 
        the optical light by the Ly$\alpha$ forest].
      Indeed, Schady et al.\ (2007) found that the X-ray afterglows 
      of GRBs not detected by UVOT were more affected by extinction 
      than those of GRBs with detected UVOT counterparts.
      The recent detection of the near infrared (IR) afterglows
      of some GRBs (which would have been considered as ``dark bursts''
      since their afterglows were not detected in any bluer bands)
      provides another piece of evidence for dust obscuration
      (e.g. see Jaunsen et al.\ 2008, Tanvir et al.\ 2008).
\item {\it Reddening} -- some GRB afterglows with low redshifts
    appear very red, due to effects of extinction -- ultraviolet
    (UV)/visible light is extinguished more by dust than red light 
    (e.g. see Klose et al.\ 2000, Levan et al.\ 2006).
    Dust reddening is also indicated by the significant deviation
    of the optical/near-IR spectral energy distributions (SEDs) of 
    many afterglows from that expected from standard models.
    Also because of dust reddening, the Balmer line ratios in 
    the spectra of some GRB host galaxies 
    (e.g. see Djorgovski et al.\ 1998),
    known as the {\it Balmer decrement},
    deviate from the expected ratios for 
    the standard Case B recombination,
    which are fairly independent of physical conditions
   (Osterbrock \& Ferland 2006).
\item {\it Depletion} -- dust-forming heavy elements 
such as Si and Fe were found to be substantially 
depleted from the gas phase in some host galaxies 
(e.g. see Savaglio et al.\ 2003). This indirectly
shows the presence of dust in GRB host galaxies
since the missing heavy elements must have
been locked up in dust grains.
\item {\it Connection between long GRBs and massive stars} 
-- there are multiple strong lines of evidence 
that long-duration ($\simgt 2\s$) GRBs 
are associated with the death of massive stars, 
occurring in regions of active star formation 
embedded in dense clouds of dust and gas
(see Woosley \& Bloom 2006).
\end{itemize}

A precise knowledge of the extinction 
(quantity, wavelength-dependence) and 
the nature (size, composition, and quantity) 
of the dust in GRB host galaxies is crucial for 
\begin{itemize}
\item Correcting for the extinction of afterglows 
    from X-ray to near-IR wavelengths to derive their
    intrinsic luminosities --
      this is particularly important
      for studying the luminosity distribution of 
      GRB afterglows and their intrinsic SEDs
      (e.g. see Kann et al.\ 2008);
\item Constraining the nature of the GRB progenitors 
    (i.e. collapsing massive stars 
    or merging neutron stars) --
      if long-duration GRBs are indeed linked to
      the collapse of massive stars, 
      it is most likely that
      their optical and near-IR afterglows will 
      suffer from significant attenuation in the 
      star-forming molecular clouds heavily 
      enshrouded by dust
      -- the birth place of these short-lived
      ($\simali 10^{6}\yrs$) massive stars;
\item Tracing the physical conditions of 
    (and processes occurring in) the environments 
    where GRBs occur which hold clues for understanding 
    the mechanism for making a burst, e.g.,
      a flat or gray extinction law for GRB host galaxies
      would imply a dense circumburst environment where 
      dust undergoes coagulational-growth 
      or a preferential destruction of small grains; and 
\item Probing the interstellar medium (ISM) 
    of high-redshift galaxies 
    and the cosmic star formation history --
      because of their intense luminosity
      which allows their detection at cosmological distances,
      GRBs are a powerful tool to study the star formation 
      history up to very high redshifts;
      e.g., the dust and extinction properties 
      of GRB hosts would help understand the nature 
      of dark bursts and the dark burst fraction
      which would place important constraints
      on the fraction of obscured star formation 
      in the universe 
      (e.g. see Djorgovski et al. 2001, Ramirez-Ruiz et al.\ 2002).
\end{itemize}

However, our current understanding of the dust extinction 
in GRB host galaxies is still very poor. 
Existing studies on this often draw conclusions
in conflict with each other 
(see \S\ref{sec:status} for details).
We argue that this could be caused by the prior
adoption of a {\it template} extinction law
in fitting the observed GRB afterglow spectra
to derive dust extinction (\S\ref{sec:status}).
We propose in this work an alternative, robust method 
based on an analytical formula which can restore the 
widely adopted template extinction laws (\S\ref{sec:approach}).
For illustration, we apply this approach to 
GRB\,000301C and GRB\,021004 (\S\ref{sec:test}).
We demonstrate in \S\ref{sec:discussion} the uniqueness 
of the derived extinction laws.
The robustness of this approach will be discussed in
a separate paper (Liang \& Li 2008a) in which the afterglow
SEDs of $>$\,50 GRBs of a wide range of properties 
are successfully modeled and for which the inferred
extinction curves are diverse, with some differing 
substantially from any of the template extinction curves. 

\section{Current Status\label{sec:status}}
At present, the amount of extinction (usually the rest-frame 
visual extinction $A_{V_r}$) and the wavelength-dependence of
the extinction (``extinction curve'' or ``extinction law''; 
$A_\lambda/A_V$ or $A_\nu/A_V$ if expressed in frequency) 
are commonly derived by fitting the UV, optical, and near-IR 
afterglow photometry 
($F_\nu$; with the Galactic extinction corrected) 
with a power-law model ($\propto \nu^{-\beta}$; 
approximating their intrinsic spectra) 
reddened by an {\it assumed}, 
template extinction law $A_\nu/A_V$
\begin{equation}
\label{eq:Fnu}
\Fnu = \Fo\,\left(\nu/{\rm Hz}\right)^{-\beta} 
\exp\left[-\frac{A_{V_r}}{1.086} \frac{A_{(1+z)\nu}}{A_{V_r}}\right]~~,
\end{equation}
where $\beta$ is the intrinsic power-law slope 
of the afterglow, 
$\Fo$ is a normalization constant 
(normalized to the overall afterglow flux level), 
$A_{(1+z)\nu}$ is the rest-frame extinction,
and $z$ is the GRB redshift.
The factor of ``1.086'' in eq.(\ref{eq:Fnu})
arises from the conversion of extinction (in magnitude)
to optical depth.
As a priori, six {\it template} extinction laws have 
been widely adopted in the literature to derive the dust 
extinction of GRB hosts:
(1) a simple power-law  
      $A_\lambda/A_V$\,$\sim$\,$\lambda^{-\gamma}$
      or even just a linear function of inverse wavelength
      $A_\lambda/A_V$\,$\sim$\,$\lambda^{-1}$
      (``Linear'' thereafter); 
(2) the Milky Way (MW) extinction curve
    (with a prominent bump at 2175$\Angstrom$)
    characterized by $R_V$, the total-to-selective
    extinction ratio (the Galactic average value is 
    $R_V\approx 3.1$);
(3) the featureless Small Magellanic Cloud (SMC) extinction curve 
     which steeply rises with inverse wavelength 
      from near-IR to far-UV
      ($A_\lambda/A_V \sim \lambda^{-1.2}$);      
(4) the Large Magellanic Cloud (LMC) curve being 
    intermediate between that of the MW and the SMC;
(5) the featureless ``Calzetti'' attenuation law 
    for the dust in local starburst galaxies 
    (Calzetti et al.\ 1994);\footnote{%
      We should note that recent Spitzer observations 
      in the near- and mid-IR argue against GRB hosts 
      being strongly starbursting galaxies
      (Le Floc'h et al.\ 2006),
      although their morphological and average radio/submillimeter 
      properties suggest that they are likely massive 
      and actively star-forming galaxies
      (Berger et al.\ 2003; Conselice et al. 2005).
      }
      and
(6) the relatively flat ``Maiolino'' extinction law for the dust 
    in the dense circumnuclear region of AGNs (Maiolino et al.\ 2001) 
    where the dust size distribution is skewed 
    toward large grains (see Fig.\,1).

To our knowledge, exceptions to the ``template''
extinction approach described here are that of 
Chen et al.\ (2006) and Li et al.\ (2008), 
both of which were based on the fireball model. 
The latter approach is limited to bursts of which 
the X-ray and optical decay indices are the same.
In most studies (which assume an extinction template)
a SMC-type extinction curve is preferred.
This is probably because the 2175$\Angstrom$ extinction
feature (which is prominent in the MW and LMC curves)  
is rarely seen in the afterglow spectra of GRBs.
So far, its possible detection is only reported in
four bursts: GRB\,970508 (Stratta et al.\ 2004),
GRB\,991216 (Kann et al.\ 2006, Vreeswijk et al.\ 2006), 
GRB\,050802 (Schady et al.\ 2007),
and more definitely GRB\,070802 
(Kr\"uhler et al.\ 2008; 
\'A. El\'iasd\'ottir et al.\ 2008, in preparation).   
   
However, some studies favour a much flatter or even gray 
extinction curve (e.g. see Savaglio et al.\ 2003, 
Savaglio \& Fall 2004, Stratta et al.\ 2005, 
Chen et al.\ 2006, Li et al.\ 2008, Perley et al.\ 2008).
With a SMC-type curve, the amount of visual extinction 
$A_V$ or reddening\footnote{%
  Reddening is usually expressed as 
  $E(B-V)\equiv A_B-A_V \equiv A_V/R_V$,
  where $A_B$ is the extinction at 
  the $B$ band ($\lambda_B\approx 4400\Angstrom$).
  By definition, gray dust (for which the extinction
  is just weakly dependent on $\lambda$) is characterized
  by small reddening $E(B-V)$ and large $R_V$.
  Apparently, for gray dust, a small {\it reddening} does not
  necessarily imply a small {\it extinction} since $R_V$ can be large.
  \label{ft:reddening}
  }
derived by fitting the afterglow photometry 
tends to be small since the SMC curve rises so rapidly 
with $\lambda^{-1}$ that a small $A_V$
would imply a large UV extinction.
This may explain the finding of ``a strong clustering 
toward low extinction ($A_V$\,$\simlt$\,0.2\,mag)''
in a detailed study of 19 GRBs by Kann et al.\ (2006),
and later by Kann et al.\ (2008) for 15 GRBs.
In contrast, for a flatter extinction law 
like that of Calzetti, Maiolino, MW with $R_V$\,$>$\,4,
or that derived by Chen et al.\ (2006), Li et al.\ (2008)
and Perley et al.\ (2008),
a relatively large $A_V$ is often obtained.

The visual extinction $A_V$ can also be 
inferred from the dust depletion method 
based on the gas-phase heavy-element abundances 
estimated from the afterglow optical absorption 
spectroscopy (Savaglio et al.\ 2003; 
Savaglio \& Fall 2004). 
  This analysis assumes both the dust depletion pattern
  and the visual extinction per unit dust column
  $A_V/N_{\rm dust}$ of GRB hosts to be the same as
  that of the MW. It is quite possible that GRB hosts may 
  have a different depletion pattern
  and/or a different $A_V/N_{\rm dust}$ conversion
  factor. The latter could result from a dust composition
  or size distribution differing from that of the MW.

One can also derive $A_V$ from
the neutral hydrogen column density $N_{\rm H}$
derived from Ly$\alpha$ absorption 
(Hjorth et al.\ 2003)
or the equivalent $N_{\rm H}$ obtained from 
soft X-ray absorption (mostly from 
oxygen K-shell absorption; Galama \& Wijers 2001,
Stratta et al.\ 2004, Watson et al.\ 2006).
There is a puzzling discrepancy between 
the optical reddening $E(B-V)$ derived from 
the afterglow SED fitting and the visual extinction
$A_V$ inferred from the dust depletion analysis
or from $N_{\rm H}$ measured from the Ly$\alpha$ 
or X-ray absorption spectra,
with the former considerably smaller than the latter.

This discrepancy problem could be alleviated
if one invokes a flat or gray extinction law.
This is because (1) gray dust ($\simgt 1\mum$) 
characterized with an extinction curve weakly 
dependent on $\lambda$ in the optical/UV  
could produce a high $A_V$ but little reddening 
(see Footnote-\ref{ft:reddening}),
and (2) per unit mass gray dust is not as effective 
as submicron-sized dust in absorbing and scattering 
optical light so that the $A_V/N_{\rm dust}$ conversion 
factor for gray dust is smaller than that of the MW dust 
(with a typical size of $\sim 0.1\mum$; see Li 2008). 
The latter would imply that the methods based on 
dust depletion (see above)
and Ly$\alpha$/X-ray absorption may overestimate $A_V$ 
if the dust size distribution of GRB host galaxies
is indeed biased toward large grains, as a result of
dust coagulational growth in the dense circumburst
environments or preferential destruction of small dust
by GRB emission (e.g. see Waxman \& Draine 2000,
Fruchter et al.\ 2001, Perna et al.\ 2003).

However, we should stress that the gray extinction hypothesis
should not be considered as the only solution to 
the discrepancy problem (after all, SMC-type or even
steeper extinction laws were derived for the hosts
of some GRBs; see Liang et al.\ 2008a).
Indeed, the mismatch between the X-ray-derived $A_V$
and that derived from the optical SED modeling
could be attributed to physically separate X-ray 
and optical emission regions (e.g. see Prochaska et al.\ 2006, 
Watson et al.\ 2007). Prochaska et al.\ (2006) argued that for 
GRB\,051111, the X-ray opacity comes from dust-free gas that is 
very local to the GRB ($\simali$1\,pc), while they placed 
a lower limit of $>$\,50\,pc on the host galaxy absorption 
systems from the GRB.
It has also been argued that $A_V$ is probably probing 
the dust outside of the dense molecular cloud around the GRB, 
since all dust within the cloud is likely to have been 
obliterated by the burst 
(e.g. see Perna \& Lazzati 2002, Prochaska et al.\ 2007).
Moreover, if the dust depletion pattern of GRB hosts is
different from that of the Milky Way, the discrepancy 
between the depletion-derived $A_V$ and that from 
the optical SED modeling could be alleviated.

\section{Our Approach\label{sec:approach}}
\begin{figure}[ht]
\begin{center}
\includegraphics[width=\figwidth,angle=0]{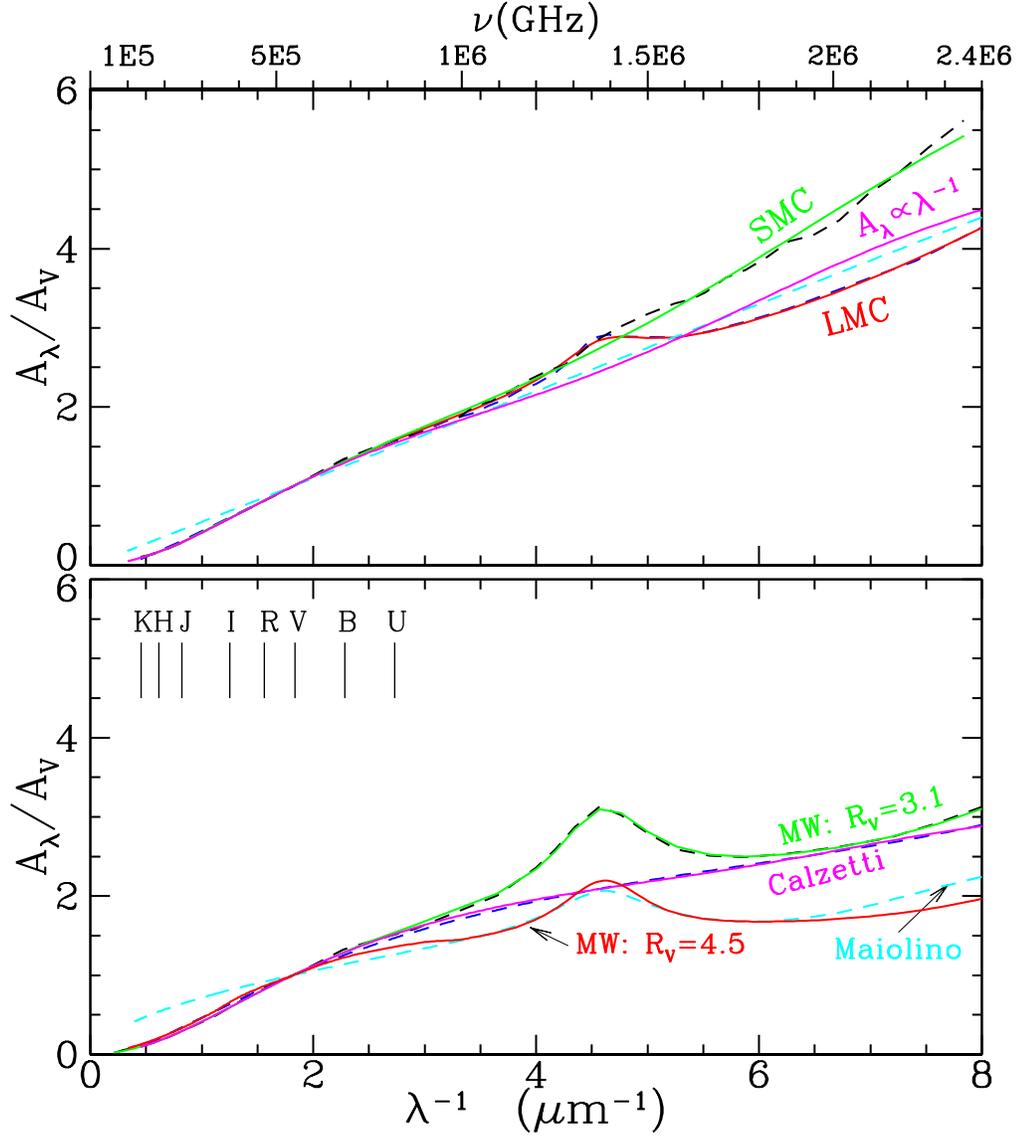}
\end{center}\vspace*{-1em}
\caption{
        \label{fig:extcurv}
        Extinction laws widely adopted 
        as {\it ``templates''} in GRB host extinction studies:
        the SMC law (upper panel: dashed black line; Pei 1992),
        the LMC law (upper panel: dashed blue line; Pei 1992), 
        the linear $A_\lambda \propto \lambda^{-1}$ law
        (upper panel: dashed cyan line),
        the MW Galactic average extinction law 
        ($R_V=3.1$; lower panel: dashed black line; Pei 1992),
        the MW extinction law with $R_V=4.5$ for dense
        clouds (lower panel: solid red line),
        the Calzetti starburst attenuation law
        (lower panel: dashed blue line),
        and the Maiolino law for AGN dust tori
        (lower panel: dashed cyan line; just like that of
         the MW with $R_V=4.5$).
        Also shown are the ``Drude'' fits to
        these ``template'' extinction laws: 
        SMC (upper panel: solid green line),
        LMC (upper panel: solid red line),
        Linear (upper panel: solid magenta line),
        MW with $R_V=3.1$ (lower panel: solid green line),
        and Calzetti (lower panel: solid magenta line). 
        }
\end{figure}

In view of the shortcomings of the prior assumption
of a template extinction law 
(see \S\ref{sec:status}) 
and guided by Pei (1992), we propose a simple formula 
containing four dimensionless parameters 
($c_1$, $c_2$, $c_3$, and $c_4$) for the wavelength-dependence 
of the extinction for
the dust in GRB host galaxies,
instead of adopting
any known extinction laws (see \S\ref{sec:status})
as a template,
\beqa
\nonumber
\label{eq:A2AV}
\Alambda/\AV & = & \frac{c_1}{\left(\lambda/0.08\right)^{c_2}
+ \left(0.08/\lambda\right)^{c_2} + c_3} \\
\nonumber
& + & \frac{233\left[1 - c_1/\left(6.88^{c_2}+0.145^{c_2}+c_3\right)
- c_4/4.60\right]}{\left(\lambda/0.046\right)^2
+ \left(0.046/\lambda\right)^2 + 90} \\
& + & \frac{c_4}{\left(\lambda/0.2175\right)^2
+ \left(0.2175/\lambda\right)^2 -1.95} ~~~,
\eeqa
where $\lambda$ is in $\mu$m.\footnote{%
  Reichart (2001) proposed a seven-parameter formula
  for the dust extinction curve $A_\lambda/A_V$ of
  GRB hosts based on the expressions of 
  Cardelli et al.\ (1989; ``CCM''; for $\lambda >0.3\mum$)
  and of Fitzpatrick \& Massa (1990; ``FM''; 
  for $0.1\mum < \lambda < 0.3\mum$).
  The problem with the Reichart (2001) formula 
  (see his eqs.\,61,66) is that the CCM expression is
  only valid for the Galactic extinction curves,
  it is not suitable for the SMC or LMC extinction 
  (Gordon et al.\ 2003).
  Therefore, if a GRB host happens to have a SMC- or LMC-type
  extinction law, models based on the Reichart (2001) formula
  will not be able to restore the true extinction.  
  } 
While the first term in the right-hand side of Eq.(\ref{eq:A2AV})
represents the far-UV extinction rise, the second term 
and the third term respectively account for the near-IR/visible
extinction and the 2175$\Angstrom$ extinction bump.
We call this the ``{\it Drude}'' approach 
since Eq.(\ref{eq:A2AV}) looks like a sum of
Drude functions.
As shown in Figure \ref{fig:extcurv}, this formula,
with the free parameters $c_j$ ($j=1,...,4$) adjusted
using the Levenberg-Marquardt minimization algorithm
(Press et al.\ 1992; see Table \ref{tab:extcurv}),
can reproduce the extinction curves 
widely adopted as template extinction laws
in GRB afterglow SED modeling,
clearly demonstrating the advantages
of the proposed formula over any template
extinction laws with a fixed wavelength-dependence shape:
with the widely-adopted conventional extinction laws 
self-contained in Eq.(\ref{eq:A2AV}) 
and the capability of revealing extinction laws 
differing from the conventional ones, 
the proposed formula is more flexible 
and more powerful in modeling the afterglow SEDs.    
Indeed, as shown in Liang \& Li (2008a,b), 
dust reddening models based on this formula nicely 
reproduce the observed afterglow SEDs of distant
GRBs at $z>4$ (including GRB\,050904 at $z\approx 6.3$) 
and that of the ``troublesome'' 
GRB\,061126 (Perley et al.\ 2008) 
without resorting to an exotic extinction law. 

\section{GRB\,000301C and GRB\,021004: Test Cases\label{sec:test}}
We apply the above-described technique (\S\ref{sec:approach})
to the optical afterglows of GRB\,000301C 
at $z\approx 2.04$ (Jensen et al.\ 2001)
and GRB\,021004 at $z\approx 2.33$ (Fynbo et al.\ 2005).
They are selected mainly because they are among 
the best-observed in terms of sampling in the time
domain, and multiwavelength coverage. 
We fit their broadband SEDs 
using Eqs.(\ref{eq:Fnu},\ref{eq:A2AV})
with $\beta$, $A_V$, $c_1$, $c_2$, and $c_3$ 
allowed to vary as free parameters [$\Fo$ is not really
a free parameter; for a given set of ($\beta$, $A_V$, 
$c_1$, $c_2$, $c_3$), $\Fo$ is uniquely determined by
the overall flux level. Therefore, in the SED modeling
we fit five free parameters to the six (seven) data points of 
GRB\,000301C (GRB\,021004)].\footnote{%
  We set $c_4$\,=\,0 based on a visual inspection
  of the observed SEDs which clearly suggest 
  the absence of a 2175$\Angstrom$ feature
  (see Figs.\,\ref{fig:GRB000301C},\ref{fig:GRB021004}).
  With $c_4$ treated as a free, positive parameter,
  even the best fits (given by $c_4\approx 0.0034,0.0018$
  for GRB\,000301C and GRB\,021004, respectively; 
  for comparison, $c_4\approx 0.051,0.039$ for
  MW and LMC, respectively) are not as good as 
  that provided by models with $c_4=0$.
  We place an upper limt of $c_4\approx 0.015$ (0.0073)
  for GRB\,000301C (GRB\,021004).
  }
We derive the best-fit parameters based on
the Levenberg-Marquardt minimization algorithm
(see Table \ref{tab:grbmod}).
As shown in Figure\,\ref{fig:GRB000301C}
for GRB\,000301C and in Figure\,\ref{fig:GRB021004}
for GRB\,021004, almost perfect fits to the observed
SEDs are achieved through this approach.
The inferred extinction curves differ substantially
from any of the template extinction laws.

\section{Discussion\label{sec:discussion}}
\begin{figure}[ht]
\begin{center}
\includegraphics[width=\figwidth,angle=0]{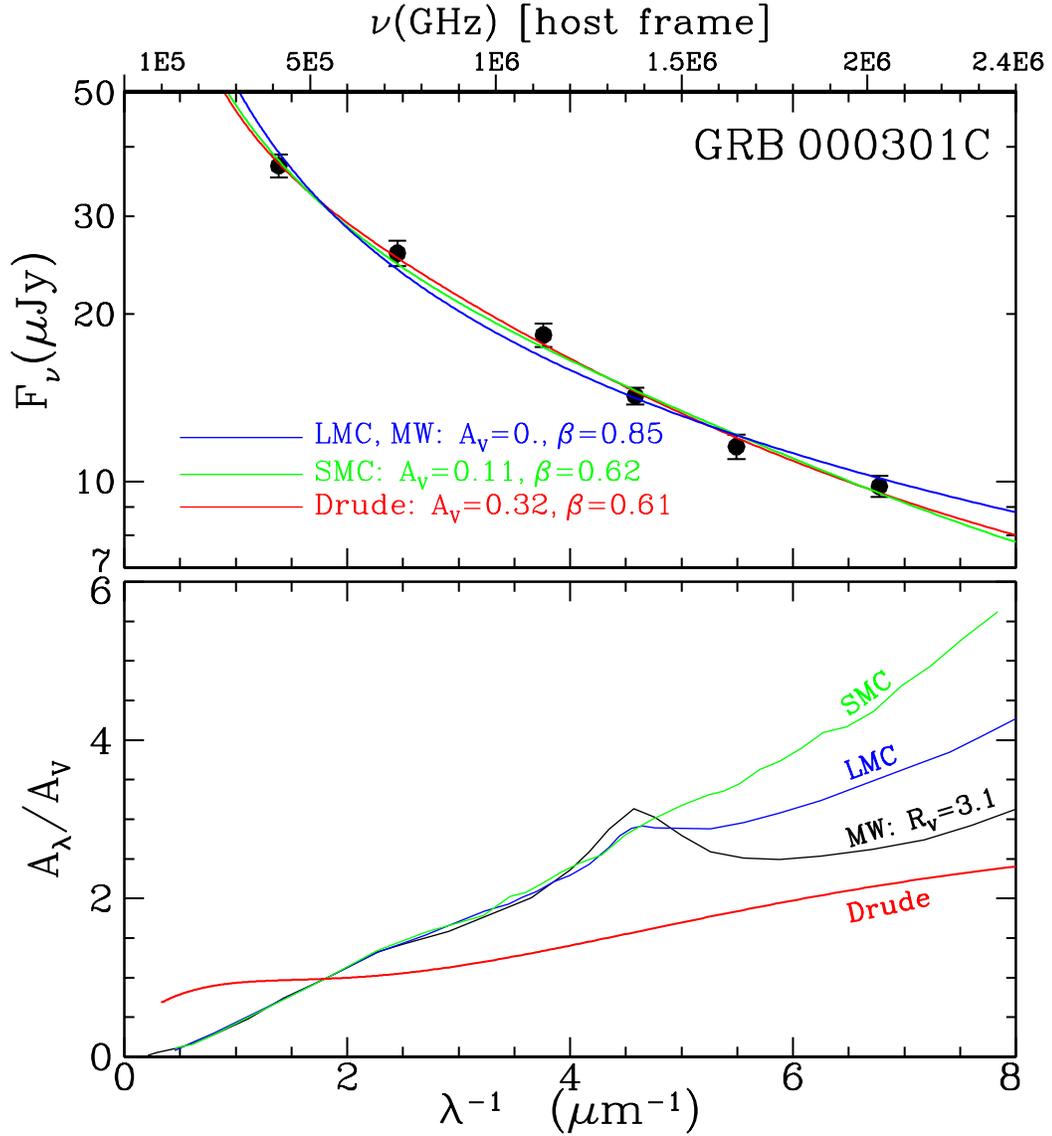}
\end{center}\vspace*{-1em}
\caption{
        \label{fig:GRB000301C}
        Upper panel: fitting the SED of the afterglow of 
        GRB\,000301C (filled black circles) 
        with the SMC (green) and LMC or MW (blue) template extinction
        laws and the ``Drude'' approach (red; see Eq.\ref{eq:A2AV})
        for the host extinction curve.
        No extinction is allowed in the MW and LMC models 
        (i.e. the best fit with a MW- or LMC-type extinction
         is given by $A_V$\,$\approx$\,0): a small amount of
        $A_V$ would lead to large deviations from the afterglow
        SED since the 2175$\Angstrom$ bump prominent in
        the MW and LMC laws is absent in the afterglow SED.
        Lower panel: comparison of the SMC (green), LMC (blue), 
        MW ($R_V=3.1$; black) extinction laws with
        that derived from the Drude approach (red).
        }
\end{figure}

\begin{figure}[ht]
\begin{center}
\includegraphics[width=\figwidth,angle=0]{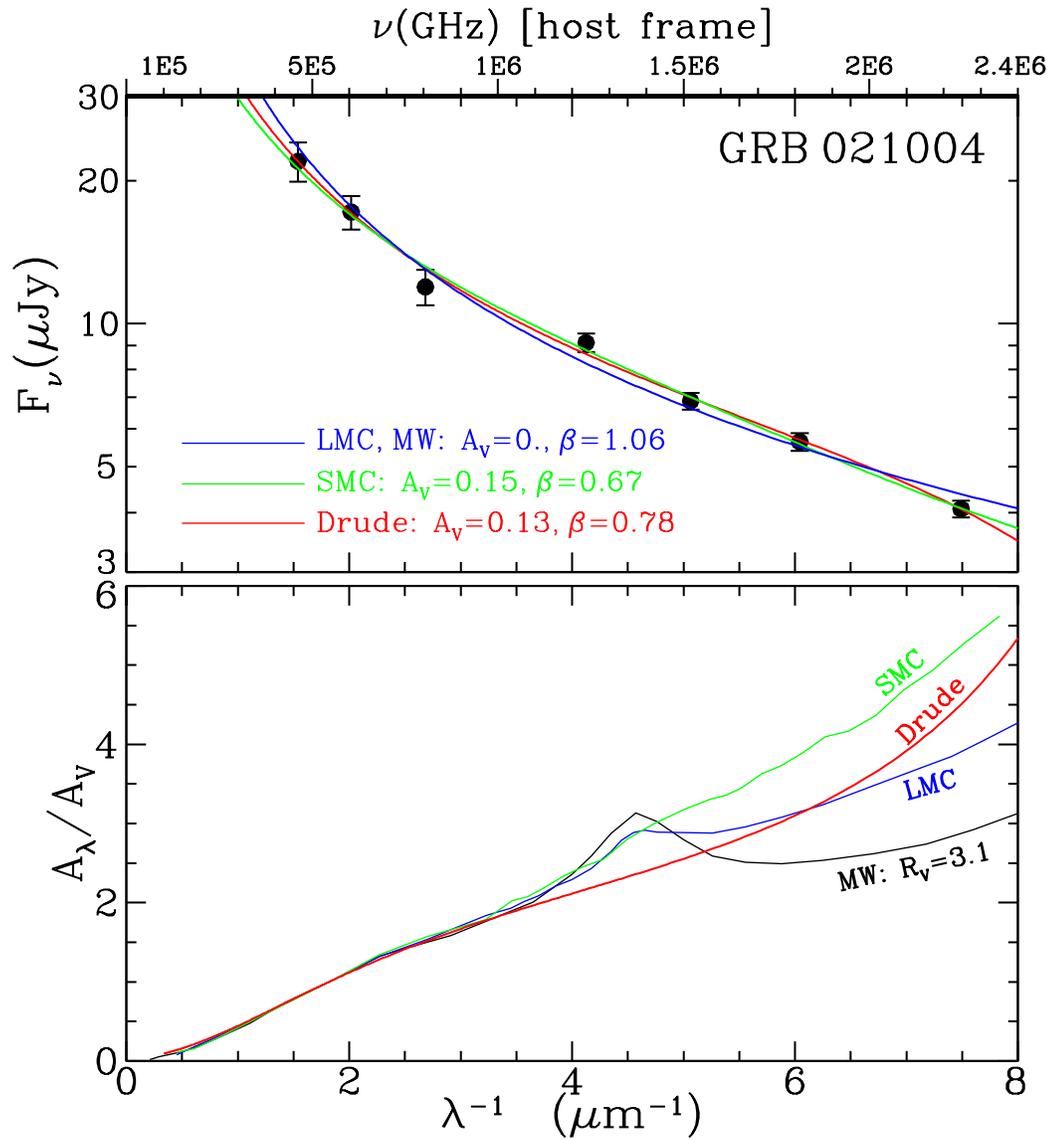}
\end{center}\vspace*{-1em}
\caption{
        \label{fig:GRB021004}
        Same as Figure \ref{fig:GRB000301C}
        but for GRB\,021004.
        }
\end{figure}

We have also fitted the afterglow
SEDs of GRB\,000301C and GRB\,021004
in terms of the MW, SMC, LMC, Calzetti,
and ``linear'' 
template extinction curves (see Table \ref{tab:grbmod} 
and Figs.\,\ref{fig:GRB000301C},\ref{fig:GRB021004}).
Since for a given template extinction law 
the wavelength-dependence of the extinction
$A_\lambda/A_V$ is fixed, we are now left with 
only three parameters: $\Fo$, $\beta$, and $A_V$.
The models based on the MW and LMC extinction laws
could not fit the observed SEDs at all. 
This is because the 2175$\Angstrom$
extinction feature which is prominent in the MW and LMC   
curves is absent in the SEDs of GRB\,000301C and GRB\,021004.
In contrast, the SMC and ``linear'' models closely fit 
the afterglow SEDs of these two bursts, 
better than the Drude model proposed here 
as measured by $\chi^2/N_{\rm d.o.f.}$
(see Table \ref{tab:grbmod}).

While the Drude model has three more parameters 
than the SMC and ``linear'' models, the quality 
of the fitting of the Drude model is even not as
good as that of the SMC or ``linear'' model. 
Then, why do not we simply adopt the SMC 
or ``linear'' model? 

First of all, we should note that there are no physical
reasons for a prior assumption of a known extinction law,
either that of the SMC, LMC, ``linear'' or MW:
the composition and size distribution
(and therefore the extinction law) of 
the dust in the dense circumburst clouds
of GRB hosts with a wide range of metallicities
and evolutionary stages 
are not expected to resemble that of 
the MW, LMC, or SMC (e.g. see Dwek 2005).
In literature, a SMC-type extinction is often assumed
for low-metallicity environments. However, there is no
physical basis for this (except the lack of grain growth
in these regions because of the lack of raw dust materials
-- the SMC dust, on average, is substantially smaller than
that of the Milky Way [see Weingartner \& Draine 2001]).
Moreover, it is known that the GRB hosts have
a wide range of metallicities. 
Indeed, the reasons why the MW, LMC and SMC 
laws are often used for GRB afterglow SED
modeling are mainly (1) little is known about the extinction 
laws of other galaxies, and (2) the Pei (1992) formula for 
the MW, LMC and SMC extinction laws is numerically convenient
for computer implementation. 

Second, although the SMC-type extinction 
is preferred in most of the present afterglow 
SED modeling studies, only the Drude approach is 
capable of reproducing the SEDs of those reddened 
by gray extinction or by non-conventional extinction.
Indeed, it was shown that the afterglow SED
of GRB\,050904 at a redshift of $z$\,$\approx$\,6.3
cannot be explained by dust reddening with 
any of the conventional (MW, SMC, Calzetti) 
extinction curves; instead, it can be well 
reproduced by invoking the extinction curve 
inferred for a distant quasar at $z$\,=\,6.2 (Maiolino et al.\ 2004),
suggesting that the properties of dust may
evolve beyond $z$\,=\,6 (Stratta et al.\ 2007).

Third, the Drude model would at least complement 
the models using template extinction curves, 
particularly for those bursts for which the Drude 
model gives a larger $\chi^2/N_{\rm d.o.f.}$ 
(but still fits the observed SEDs well).  
Given that the derived extinction $A_V$ 
and the intrinsic spectral slope $\beta$ 
differ appreciably among different approaches
(see Table \ref{tab:grbmod}), the SMC model
(and other models) should be used along side
with the Drude model to gain insight into
the ``true'' extinction and the ``true'' spectral slope.

\begin{figure}[ht]
\begin{center}
\includegraphics[width=\figwidth,angle=0]{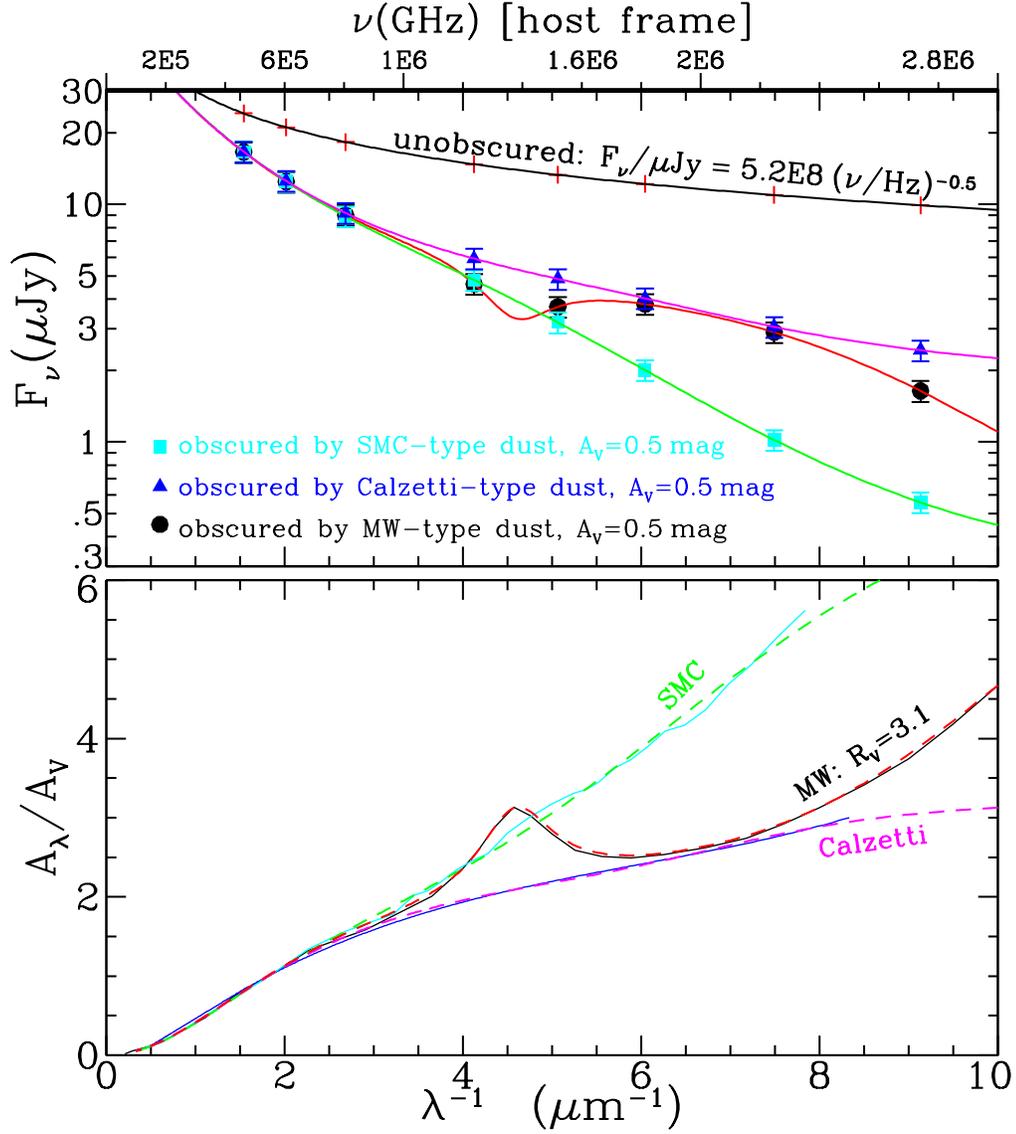}
\end{center}\vspace*{-1em}
\caption{
        \label{fig:test} 
        Upper panel: Drude fits to the observer-frame
        $UBVRIJHK$ ``photometry data''
        artificially-generated by reddening the intrinsic afterglow
        spectrum $F_\nu\propto \nu^{-\beta}$ of 
        a burst at $z\approx 2$
        (black line with red crosses superimposed for 
         the observer-frame $UBVRIJHK$ bands) with 
        the SMC (``data'': cyan squares; Drude fit: green line),
        Calzetti (``data'': blue triangles; Drude fit: magenta line),
        and MW (``data'': black circles; Drude fit: red line)
        extinction laws (with $A_V=0.5\magni$ for each).
        Lower panel: comparison of the SMC, Calzetti and MW
        extinction curves (solid lines) with that inferred from 
        the Drude approach (dashed lines).
        }
\end{figure}

We finally demonstrate the uniqueness of the extinction
curve inferred from the Drude approach. 
To this end, we generate three sets of afterglow
``photometry data'' by reddening the intrinsic afterglow 
spectrum $F_\nu\,(\mu {\rm Jy})$\,=\,$5.2\times 10^8 
\left(\nu/{\rm Hz}\right)^{-0.5}$ of a burst at $z\approx 2$ 
respectively with three template extinction laws: 
MW, SMC, and Calzetti, each with $A_V=0.5\magni$. 
We then apply the Drude approach to these three sets 
of artificially-created GRB afterglow data. 
As shown in Figure\,\ref{fig:test},
we uniquely restore the MW, SMC, and Calzetti extinction
laws: the inferred extinction curves are almost identical
to that used to redden the intrinsic spectrum
(the derived parameters [see Table\,\ref{tab:test}]
are essentially the same as those tabulated 
in Table\,\ref{tab:extcurv}).

\acknowledgments
We thank the anonymous referee and J.X. Prochaska 
for very helpful comments.
A.L. and S.L. are supported in part by a NASA/Swift Theory Program, 
a NASA/Chandra Theory Program, and the NSFC Outstanding Oversea 
Young Scholarship. D.M.W. is supported by the NSFC grants
10621303 and 10673034, and the National Basic Research Program of
China (973 Program 2007CB815404). D.A.K. and S.K. acknowledge 
financial support by DFG grant Kl 766/13-2.

\begin{table}
{\footnotesize
\caption[]{\footnotesize
           ``Drude'' fits to known extinction curves
           for $\lambda$\,=\,0.1--1$\mum$
           widely adopted as ``templates'' in modeling 
           GRB afterglow SEDs to derive GRB host dust extinction.
           \label{tab:extcurv}}
\begin{center}
\begin{tabular}{ccccccc}
\tableline\tableline
 & $c_1$ & $c_2$ & $c_3$ & $c_4$ & $\chi^2/\dof$ \\
\cline{1-6}
MW       & 14.4 & 6.52  & 2.04  & 0.0519 & 1.66 \\
LMC      & 4.47 & 2.39  & -0.988 & 0.0221 & 1.19 \\
SMC      & 38.7 & 3.83  & 6.34 & 0. & 1.36 \\
Linear   & 66.2 & 4.97  & 22.1 & 0. & 1.42 \\
Calzetti & 44.9 & 7.56  & 61.2 & 0. & 1.68 \\
\tableline
\end{tabular}
\end{center}
}
\end{table}

\begin{table}
{\footnotesize
\caption[]{\footnotesize
           Results of fitting to the afterglow SEDs
           of GRB\,000301C and GRB\,021004 
           with the Drude approach 
           (see \S\ref{sec:approach}, \S\ref{sec:test})
           or various template extinction laws.
           Note that the Drude approach has more free
           parameters than the other approaches.
           \label{tab:grbmod}}
\begin{center}
\begin{tabular}{ccccccccccc}
\tableline\tableline
Extinction & $c_1$ & $c_2$ & $c_3$ & $c_4$  
           & $\AV$ & $\beta$ & $\Fo$     
           & $\chi^2/N_{\rm data}$ 
           & $\chi^2/N_{\rm d.o.f.}$ \\
Type       &  &  &   &   & (mag) &         & ($\mu$Jy) &    \\  
\cline{1-10}
&       &       &       &  &  GRB\,000301C & &  & &  &               \\ 
\cline{1-10} 
Drude &0.025 &0.048 &-2.00  & 0. &0.32 &0.61  &3.99E10 & 0.33 & 1.98\\
MW    & ...   & ...   & ...   & ... &0. & 0.85 &1.04E14 & 1.32 &2.64\\
SMC   & ...   & ...   & ...   & ... &0.11  & 0.62  &4.68E10  & 0.64 &1.28\\
LMC   & ...   & ...   & ...   & ... &0. & 0.85 &1.04E14  & 1.32 &2.64\\
Linear   & ... & ...  & ...   & ... &0.20  & 0.51  &1.23E9  & 0.47 &0.94\\
Calzetti & ... & ...  & ...   & ... &0. & 0.85 & 1.04E14 & 1.32 &2.64\\
\cline{1-10}
&       &       &       &   & GRB\,021004   & &  & &  &               \\ 
\cline{1-10}
Drude &0.015 &0.15 &-2.00  & 0. &0.13 &0.78  &6.13E12 &0.47 &1.64\\
MW    & ...   & ...   & ...   & ... &0.  & 1.06    &8.23E16  & 1.53 &2.68\\
SMC   & ...   & ...   & ...   & ... &0.15  & 0.67  &1.58E11   & 0.36 &0.53\\
LMC   & ...   & ...   & ...   & ... &0.  & 1.06  &8.23E16  & 1.53 &2.68\\
Linear   & ... & ...  & ...   & ... &0.26  & 0.54  &2.05E9   & 0.76 &1.33\\
Calzetti & ... & ...  & ...   & ... &0.95  & 0.    &46.7     & 0.80 &1.40\\
\tableline
\end{tabular}
\end{center}
}
\end{table}

\begin{table}
{\footnotesize
\caption[]{\footnotesize
           Results of Drude-fitting to the artificial SED
           generated by reddening the power-law afterglow           
           $F_\nu\,(\mu {\rm Jy})$\,=\,$5.2\times 10^8 
           \left(\nu/{\rm Hz}\right)^{-0.5}$ 
           with $A_V$\,=\,0.5\,mag extinction 
           of MW, SMC, and Calzetti-type
           (see Fig.\,\ref{fig:test}).
           \label{tab:test}}
\begin{center}
\begin{tabular}{ccccccccccc}
\tableline\tableline
Reddening & $c_1$ & $c_2$ & $c_3$ & $c_4$  
           & $\AV$ & $\beta$ & $\Fo$     
           & $\chi^2/N_{\rm data}$ 
           & $\chi^2/N_{\rm d.o.f.}$ \\
Type       &  &  &   &   & (mag) &         & ($\mu$Jy) &    \\  
\cline{1-10} 
MW    & 14.3 & 6.49  & 2.02   & 0.0514
      &0.501 & 0.499 &5.24E8 & 3.26E-4 &4.35E-4\\
SMC   & 39.4 & 3.89  & 6.31  & 0.
      &0.500 & 0.501 &5.26E8 & 1.32E-3 &1.76E-3\\
Calzetti  & 45.2 & 7.51  & 61.7  & 0.
      &0.497 & 0.502 &5.17E8 & 7.98E-4 &1.06E-3\\
\cline{1-10}
\tableline
\end{tabular}
\end{center}
}
\end{table}


\end{document}